\documentclass[12pt]{article}
\usepackage{latexsym,amssymb,epsfig}
\newcommand{\eqn}[1]{(\ref{#1})}
\newcommand{\eq}[1]{Eq.~(\ref{#1})}
\newcommand{\secn}[1]{Section~\ref{#1}}
\newcommand{\sect}[1]{\section{#1}}
\newcommand{\bra}[1]{\langle{#1}|}
\newcommand{\ket}[1]{|{#1}\rangle}
\newcommand{\dket}[1]{|{#1}\rangle\hskip -3pt\rangle}

\def\beq{\begin{equation}}
\def\eeq{\end{equation}}
\newcommand{\bea}{\begin{eqnarray}}
\newcommand{\eea}{\end{eqnarray}}
\newcommand{\Z}{\mathbb{Z}}

\def\ii{{\rm i}}
\def\ee{\mathrm{e}}

\def\Tr{{\rm Tr}}
\def\a{\alpha}

\def\k{\kappa}                  

\def\m{\mu}
\def\n{\nu}

\begin{document}
\def\thefootnote{\alph{footnote}}
\vskip0.2cm
\begin{flushright}
DFTT 14/2001\\
\end{flushright}
\vskip0.5cm
\begin{center}
{\Large\bf 
Classical geometry and gauge duals for fractional branes on
ALE orbifolds
}
\end{center}
\vskip 0.5cm
\centerline{
M. Bill\'o, L. Gallot, A. Liccardo%
\footnote{e--mail: {\tt billo, gallot, liccardo@to.infn.it}}}
\vskip0.4cm
\centerline{\sl Dipartimento di Fisica
Teorica dell'Universit\`a di Torino and}
\centerline{\sl Istituto Nazionale di Fisica Nucleare, Sezione di Torino}
\centerline{\sl via P.Giuria 1, I-10125 Torino, Italy}
\vskip 0.5cm
\begin{abstract}
We investigate the classical geometry corresponding to a collection  of
fractional D3 branes in  the orbifold limit  $\mathbb{C}^2/\Gamma$ of an ALE
space. We discuss its interpretation in terms of the world-volume gauge theory
on the branes, which is in general a non conformal $\mathcal{N}=2$ Yang-Mills
theory with matter. The twisted fields reproduce the perturbative behaviour of
the gauge theory. We regulate the IR singularities for both twisted and
untwisted  fields by means of an enhan\c con mechanism qualitatively
consistent  with the gauge theory expectations. The five-form flux decreases
logarithmically towards the IR with a coefficient dictated by the gauge theory
$\beta$-functions. 
\end{abstract}
\setcounter{footnote}{0}
\def\thefootnote{\arabic{footnote}}
\sect{Introduction}
The study of fractional branes, that is of branes wrapped on vanishing cycles
at orbifold or conifold singularities, has attracted in the last period much
interest \cite{Klebanov:2000rd}-\cite{Petrini:2001fk}. The resulting
world-volume gauge theories, in fact, possess fewer supersymmetries than the
theories on unwrapped (or ``bulk'') branes, and in  general are not conformal.
At the same time, it is often possible to derive explicit supergravity
solutions describing the geometry created by collections of such wrapped
branes.  This is therefore an arena where one can try to extend the Maldacena
gauge/gravity duality \cite{mal} to non-conformal cases, an important goal in
present day string theory \cite{tipo0}-\cite{CVETIC}.  Although the
supergravity solutions representing non-conformal situations  typically exhibit
IR singularities, it is hoped that such singularities are resolved or explained
by some stringy phenomenon.  Interesting physics has been elucidated  by
considering D3-branes on conifold singularities 
\cite{Klebanov:2000rd}-\cite{TSEYTLIN1}, corresponding to $\mathcal{N}=1$ gauge
theories, for instance the cascade of Seiberg dualities of \cite{KLEBA3}. We
will focus on the case of fractional D3-branes solutions 
\cite{Klebanov:2000rd},\cite{Bertolini:2000dk}-\cite{Petrini:2001fk} on
orbifold backgrounds of the type  $\mathbb{R}^{1,5}\times
\mathbb{C}^2/\Gamma$,  where $\Gamma$ is a Kleinian subgroup of SU$(2)$. These
fractional branes can be interpreted as D5-branes wrapped on holomorphic
two-cycles  of an ALE manifold  \cite{Douglas:1997xg}, in the limit in which
the volumes of such cycles vanish and the ALE space degenerates to
$\mathbb{C}^2/\Gamma$.  The world-volume theories are $\mathcal{N}=2$
super-Yang-Mills theories in 4 dimensions, with, in general, matter content
(hyper-multiplets). In the $\mathcal{N}=2$ context, the resolution of the IR
singularities has been often attributed to the so-called enhan\c con mechanism
\cite{enhanc}, and the fractional branes on $\mathbb{C}^2/\mathbb{Z}_2$ 
orbifolds seem to follow this pattern 
\cite{Bertolini:2000dk,Polchinski:2000mx,Petrini:2001fk}.
\par
To search for a generalization of the Maldacena conjecture in the context of
fractional branes on orbifold singularities, a possible approach  is to start
with a situation  involving only ``bulk'' (or ``regular'')  branes.  On such
branes lives a conformal field theory \cite{Douglas:1996sw,Johnson:1997py},
dual to  $\mathrm{AdS}_5\times S^5/\Gamma$ supergravity \cite{conforb}.  One
can then break conformal invariance by properly decomposing the bulk branes in
their fractional branes constituents, and placing some of the latter at a large
distance $\rho_0$ in the transverse $z$-plane  ($z=x^4+\ii x^5$) fixed by the 
orbifold action \cite{Polchinski:2000mx,Aharony,Petrini:2001fk}. This allows to
discuss the gravity dual of the non-conformal theory living on the remaining
fractional branes (of which $\rho_0$ represents the UV cut-off), as a
deformation of  the  $\mathrm{AdS}_5\times S^5/\Gamma$ geometry
\cite{Klebanov:2000rd,Petrini:2001fk}. To describe a direct gauge/gravity
duality for  a generic  configuration of (many) fractional branes, one should,
in the above perspective, send the UV cut-off  $\rho_0$ to infinity so that
the  gauge theory is valid at  all scales.  It is argued in
\cite{Aharony,Petrini:2001fk}  that it is  impossible to do so  while retaining
the supergravity approximation,  which requires large numbers of branes, as it
is generally to be  expected in presence of UV free gauge factors. 
\par
Even with this limitation, several papers have investigated the case
$\mathbb{C}^2/\mathbb{Z}_2$, determining 
\cite{Bertolini:2000dk,Polchinski:2000mx} the classical solutions%
\footnote{The classical solutions for fractional branes on 
$\mathbb{C}^2/\mathbb{Z}_N$ case were investigated, with a different approach 
from ours, in \cite{Rajaraman:2000dn}. The classical solutions for the compact 
orbifold $T^4/\mathbb{Z}_2$ have been investigated in 
\cite{anto,Merlatti:2001ne}}  and pointing out many intriguing features in
their interpretation  \cite{Polchinski:2000mx,Aharony,Petrini:2001fk}. Here we
will work out the  supergravity solutions describing fractional  branes in the
case of a generic orbifold $\mathbb{C}^2/\Gamma$ and discuss their field theory
interpretation.   One of the  main features of these solutions (which already
emerged in  the $\mathbb{Z}_2$ case)  is the existence of  non-trivial
``twisted''  complex scalar fields $\gamma_i$,  which exhibits a logarithmic
behaviour  that matches the one of the (complexified) running coupling
constants of the world-volume gauge theories. This perfect correspondence at
the perturbative level is basically due to the open/closed string duality of
the one-loop cylinder diagrams that, in the   field theory limit of the open
string channel  represent, because of the $\mathcal{N}=2$ supersymmetry, the
only perturbative corrections to the gauge theory.  
\par
In working out the solution, however, one encounters IR singularities in both
the twisted and untwisted equations of motion which need the introduction of
appropriate regulators  $\Lambda_i$.  On the gauge side these IR cut-offs can 
naturally be interpreted as the  dynamically generated scales of the various
factors of the  gauge theory,  while  on the gravity side they represent 
enhan\c con radii at which fractional probes of type $i$  become tensionless%
\footnote{This generalizes the case of $\mathbb{C}^2/\mathbb{Z}_2$, where there
are only two types of fractional branes and a single enhan\c con radius.}. From
the gauge theory point of view, drastic  modifications  with respect to the
logarithmic behaviour of the effective couplings occur at the scales
$\Lambda_i$. These modifications are due  to  the instanton corrections  and
should correspond,  on the string theory side, to fractional D-instanton
corrections  \cite{Klebanov:2000rd}. For large number of fractional branes the
instanton corrections are well implemented by assuming that there is no running
of  each gauge coupling below the corresponding scale $\Lambda_i$ \cite
{Petrini:2001fk}.  This way of regulating the  IR singularities suggests that
the fractional branes that represent the source for the supergravity solution, 
initially thought of as being placed at the origin, dispose  in fact
themselves  on (or near to) the  enhan\c con radii used as cut-offs.
\par
In the classical supergravity solution the twisted fields back-react, 
providing a source in the bulk for the untwisted fields (metric and RR 5-form).
An intriguing role  seems to be played by the RR five-form flux, which measures
the total untwisted D3-charge within a certain scale.  This flux varies
logarithmically  with the radial distance in the $z$--plane
\cite{Polchinski:2000mx,Aharony}  Such a behaviour is quite general for
fractional brane solutions in different contexts \cite{KLEBA3} and its
interpretation at the level of the field theory  is an important issue. The
flux is expected to be related to the degrees of freedom of the gauge theory;
the fact that it turns  out to decrease with the scale towards the IR is
consistent with this expectation. The untwisted charge of the branes of each
type is proportional to $\gamma_i(z)$. At the enchan\c con scale $\Lambda_i$ it
changes sign, so that below $\Lambda_i$ the system is no longer BPS.  Here a
difference with respect to the  $\mathbb{C}^2/\mathbb{Z}_2$ case  seems to
occur. Indeed in the $\mathbb{C}^2/\mathbb{Z}_2$ orbifold one can  revert to a
BPS situation by a shift of  $\gamma$ corresponding to a shift of the $B$-flux
under which the string theory background is periodic. However, the five-form
flux is not invariant unless one decreases at the same time the  number of bulk
branes by an appropriate amount. In particular starting with a configuration of
$N+M$ branes of one type and $N$ of the other, so that the world-volume theory
is a $\mathrm{SU}(N+M)\times  \mathrm{SU}(N)$ gauge  theory with matter, one
needs to add $M$ bulk branes modifying the gauge group to 
$\mathrm{SU}(N)\times \mathrm{SU}(N-M)$. This property has been interpreted
\cite{Polchinski:2000mx} as suggesting an $\mathcal{N}=2$  analogue of the
duality cascades that take place \cite{KLEBA3} in the  $\mathcal{N}=1$
theories  associated to fractional branes at the conifold. In \cite{Aharony}
the reduction of the gauge group has been instead interpreted as due to a Higgs
phenomenon.  In both interpretations, the behaviour of the untwisted fields is
taken to suggest a possible ``extension'' of the validity of the description
beyond the perturbative regime of the original  gauge theory, namely below the
dynamically generated scale. The authors  of \cite{Petrini:2001fk} take a more
conservative point of view. Relying on the comparison with the appropriate 
Seiberg-Witten curve which they construct explicitly, they argue  that it is
not  necessary to try to extend the supergravity description below the  
enhan\c con.
\par
By considering fractional branes on $\mathbb{C}^2/\Gamma$ orbifolds, with gauge
groups consisting in general of more than two factors, we find that it is no
longer true that all non-BPS situations, where the untwisted charges of the
various types do not have all the same sign, can be amended by shifts of the
twisted fields corresponding to periodicities of the $B$-fluxes. A duality
cascade seems therefore unlikely, at least if it should generalize directly the
proposal of \cite{Polchinski:2000mx}, and we tend to the ``conservative'' point
of view that limits the validity of the classical supergravity solution to the
perturbative regime of the field theory. Also remaining within this regime, the
role of the RR flux could indeed be that of counting the  degrees of freedom.
To this purpose, we note that the logarithmic running of  the flux corresponds
to a differential equation with a strong analogy (totally at a formal level,
though, at least for now)  with the equation satisfied by the holographic
$c$-function proposed in  \cite{Anselmi:2000fu}. 
\par
The paper is organized as follows. In section 2, we review some needed material
about the $\mathbb{C}^2/\Gamma$-type orbifold of type IIB, the corresponding
fractional D3-branes and their world-volume gauge theory.  In section 3, we
write the corresponding supergravity equation and solve them  using an ansatz
similar to \cite{Bertolini:2000dk,anto}. In section 4, we  discuss the relation
of the supergravity solution to the gauge theory. In appendix \ref{app:a}, we
give an analytic study of the function governing the untwisted fields in the
supergravity solution, while in appendix \ref{app:b} we review some material
about closed strings on $\mathbb{C}^2/\Gamma$ orbifolds and the boundary states
for the corresponding branes.
\sect{Fractional D3-branes on $\mathbb{C}^2/\Gamma$}
\label{sec:frac}
Let us consider the bulk theory of  type IIB strings on  
$\mathbb{R}^{1,5}\times \mathbb{C}^2/\Gamma$, where $\Gamma$ is a discrete
subgroup of SU$(2)$ acting on the coordinates $z^1\equiv x^6 + \ii x^7$ and
$z^2\equiv x^8 + \ii x^9$ by
\beq
\label{Gammaaction}
g\in\Gamma: ~{z^1 \choose z^2} \mapsto \mathcal{Q}(g){z^1 \choose z^2}~,
\eeq   
with $\mathcal{Q}(g)\in \mathrm{SU}(2)$ being the representative of $g$ in the
defining two-dimensional representation.
\par
This space admits obviously an exact description as a $\mathcal{N}=(4,4)$ SCFT,
see, \emph{e.g.},~\cite{Anselmi:1994sm}. The massless spectrum, in particular,
will be briefly described in Appendix \ref{app:b}. In the untwisted NS-NS and
R-R sectors the  $\Gamma$-invariant subset of the usual type IIB fields are
kept. These fields may depend on the whole 10-dimensional space. The twisted
fields, instead, only have a six-dimensional dynamics, as they cannot carry
closed string momentum  in the orbifolded directions. In the NS-NS twisted
sectors one has 4 six-dimensional scalars. From the R-R twisted sector, we get
a 0-form and a 2-form (always in the six-dimensional sense).
\par
This exact background, preserving half of the bulk supersymmetries, is the
singular limit of an (almost) geometrical background $\mathbb{R}^{1,5}\times
\mathcal{M}_\Gamma$, where  $\mathcal{M}_\Gamma$ is the ALE space obtained
resolving the $\mathbb{C}^2/\Gamma$ orbifold~\cite{kronheimer}. Let us recall 
that the resolved space acquires a non-trivial homology lattice
$H_2(\mathcal{M}_\Gamma,\mathbb{Z})$, entirely built by exceptional divisors.
These latter are holomorphic cycles $e_i$,  which topologically are spheres.
They are in one--to--one correspondence with the simple roots $\alpha_i$ of a
simply-laced Lie algebra $\mathcal{G}_\Gamma$ which is associated to $\Gamma$
by the McKay  correspondence \cite{mckay}. This correspondence can be stated as
follows. Denote by $\mathcal{D}_I$ the irreducible representations of $\Gamma$.
Then the Clebsh-Gordon coefficients $\widehat A_{IJ}$ in the decomposition
\beq
\label{McKay}
\mathcal{Q}\otimes \mathcal{D}_I = \oplus_J \widehat A_{IJ}\, \mathcal{D}_J
\eeq
represent the adjacency matrix of the \emph{extended} Dynkin diagram of
$\mathcal{G}_\Gamma$. The intersection matrix of the exceptional cycles is  the
negative of the Cartan matrix of $\mathcal{G}_\Gamma$:
\beq
\label{intersection}
e_i\cdot e_j = -C_{ij}~.
\eeq
The trivial representation $\mathcal{D}_0$ of $\Gamma$, on the other hand, is
associated  to the extra site in the Dynkin diagram, corresponding to the
lowest  root $\alpha_0 = -\sum_i d_i\alpha_i$, and thus to the 
(non-independent) homology cycle $e_0 = -\sum_i d_i e_i$. We will always use
the convention that the index $I$ runs on the irreducible  representations of
$\Gamma$, while $i$ runs on the holomorphic cycles  $e_i$; so, $I=(0,i)$. The
Dynkin labels $d_i$ give the dimensions of the irreducible representations
$\mathcal{D}_i$; of course, $d_0=1$.
\par
Let us notice that the orbifold CFT of $\mathbb{C}^2/\Gamma$ is obtained in the
limit in which the volume of the exceptional cycles  vanishes, but the
integrals of the $B$-field on these cycles remain finite \cite{bflux}, with the
values 
\beq
\label{fbs1}
b_i \equiv {1\over 2\pi} \int_{e_i} B = {d_i\over\Gamma}~.
\eeq
We are using here and in the following the convention that $2\pi\alpha'=1$, 
otherwise in \eq{fbs1} the string tension $1/(2\pi\alpha')$ should further
appear in front of the integral.  The limits in which also the $B$-fluxes
vanish correspond to more exotic phases in which tensionless strings appear
(and one has an effective six-dimensional theory which includes so-called
``little strings') \cite{littlestrings}.  At the string theory level, the
$B$-fluxes $b_i$ are periodic variables of  period $1$; indeed, the
contribution to the partition function of a world-sheet instanton on the
vanishing cycle $e_i$ is $\exp (2\pi\ii b_i)$.  
\par
Our purpose is to discuss the effect of placing in this background D3-branes
transverse to $\mathbb{C}^2/\Gamma$. In \cite{Douglas:1996sw} (see also
\cite{Johnson:1997py}) the D$p$-branes associated to Chan-Paton factors
transforming in the regular representation of $\Gamma$ were considered. Such
D$p$-branes are called ``bulk'' branes, as they can move in the orbifold space,
and are just the counterparts of the usual D$p$-branes of flat space.
\par
The regular representation $\mathcal{R}$ is not irreducible: it  decomposes as
$\mathcal{R}=\oplus_I d_I \mathcal{D}_I$, where $d_I$ denotes the dimension of
$\mathcal{D}_I$. It is natural to consider ``fractional'' branes
\cite{Douglas:1997xg} corresponding to Chan-Paton factors transforming in
irreducible representations. The McKay correspondence associates the
irreducible representations $D_i$ ($i\not=0$) to the homology cycles%
\footnote{And the trivial representation $D_0$ to cycle $e_0 = -\sum_i d_i
e_i$, corresponding to the lowest root of $\mathcal{G}_\Gamma$, \emph{i.e.}, to
the extra dot in the extended Dynkin diagram.} $e_i$. This points to a
geometrical explanation of such branes~\cite{Diaconescu:1998br}:   a
D$(p+2)$-brane of the theory defined on the ALE space $\mathcal{M}_\Gamma$
which wraps the homology cycle $e_i$ gives rise in the orbifold limit to the
fractional D$p$-brane%
\footnote{The fractional D$p$-brane associated to the trivial representation
$\mathcal{D}_0$ is obtained by wrapping a D$(p+2)$-brane on the cycle $e_0 =
-\sum_i d_i e_i$, but with an additional background flux of the world-volume 
gauge field $\mathcal{F}$ turned on: $\int_{e_0} \mathcal{F} = 2\pi$. This
ensures that such brane gets an untwisted D$p$-brane charge of the same sign of
those of the branes associated to non-trivial representations.} of type $D_i$.
Notice that the fractional D3 branes, stretching along the $x^0,\ldots, x^3$
directions, are confined to the plane $(x^4,x^5)$ fixed by the orbifold action
(it is only the regular representation which contains all the images of a
brane, not the single irreducible representations).
\par
The massless spectrum of the open strings stretched between the fractional 
branes determines the field content of the gauge field theory living on their
world-volumes~\cite{Douglas:1996sw}.  The spectrum is easily obtained  by
explicitly computing the  $\Gamma$-projected 1-loop trace 
\beq
\label{mk1}
Z_{IJ}(q) = 
{1\over |\Gamma|} \sum_{g\in \Gamma} \Tr_{IJ}(\hat g\, q^{L_0 - {c\over
24}})~,
\eeq     
where $\hat g$ acts on the string fields $X^\mu$ and $\psi^\mu$, but also on
the Chan-Paton labels at the two open string endpoints, transforming
respectively in the representation $\bar\mathcal{D}_I$ and $\mathcal{D}_J$. 
The trace in \eq{mk1} is of course composed of GSO-projected NS and R 
contributions. The final result is that, with $m^I$ D3-branes of each type $I$,
the massless fields are those of an $\mathcal{N}=2$ gauge theory. The gauge
group is $\otimes_I\, U(m^I)$, each  $U(m^I)$ gauge multiplet coming from the 
strings attached to the branes of type $I$, and there are $\widehat A_{JK}$
hyper-multiplets in the bi-fundamental  representation $(m_I,m_J)$ of two gauge
groups U$(m_I)$ and U$(m_J)$. These latter arise from  strings  stretched
between branes of type $I$ and $J$ whenever the link $IJ$ is in the extended
Dynkin of $\mathcal{G}_\Gamma$ (namely, $\widehat A_{JK}$ is the adjacency 
matrix of the    diagram). Let us notice that the 1-loop coefficients of the
$\beta$-functions of these gauge theories (which are also, because of the
$\mathcal{N}=2$  supersymmetry, the only perturbative contributions), are 
simply expressed in terms of the extended Cartan matrix as
\beq
\label{fbs2}
2 m_I - \widehat A_{IJ}\, m^J = \widehat C_{IJ}\, m^J~,
\eeq 
the positive contributions being of course from the non-abelian gauge multiplet
fields and the negative ones from the charged hyper-multiplets.
\par
A bulk brane corresponds, by decomposing the regular representation,  to a
collection of fractional branes with $m^I = d^I$, and, in accordance with
\eq{fbs2}, it is conformal, as the vector $d^I$ of the dimensions is the null
eigenvector of the extended Cartan matrix.  Being conformal, the field theory
associated to $N$ bulk branes is dual,  for large $N$, to type IIB supergravity
on $\mathrm{AdS}_5\times S^5/\Gamma$  \cite{conforb}. On the other hand,  the
moduli space of the field theory, beside the Higgs branch corresponding  to the
possible motion of  the bulk branes inside $\mathbb{C}^2/\Gamma$,  has a
Coulomb branch in which the  expectation values of the complex scalars  sitting
in the gauge multiplets represent the positions of the fractional  branes into
which the bulk branes can decompose when hitting the fixed plane.   To study
generic, non conformal, configurations of  fractional branes, one can start
with a stack of bulk branes and then place some   of the fractional branes
arising from their decomposition  at a large distance $\rho_0$ in the fixed
plane \cite{Polchinski:2000mx,Aharony,Petrini:2001fk}. For scales below 
$\rho_0$, the gauge theory is effectively the one of the remaining fractional
branes. This  gives the possibility of discussing the gravity dual of this
latter, non-conformal, theory as a deformation of the $\mathrm{AdS}_5\times
S^5/\Gamma$ geometry induced by the non-trivial v.e.v's.     
\sect{Classical supergravity solution}
\label{soluz}
As we discussed in the introduction, it is already clear from the
$\mathbb{C}^2/\mathbb{Z}_2$ case that the supergravity solution created by a
generic collection of fractional branes does not lead straight--forwardly to a
gauge/gravity correspondence, because sending the UV cutoff $\rho_0$ of the
gauge theory to infinity while keeping finite the dynamically  generated scale
$\Lambda$, in order to decouple it, is inconsistent with the validity of the
supergravity approximation which requires a large number of branes.  It is
nevertheless possible to derive the classical supergravity solution  that 
describes the classical geometry created by a generic configuration of 
fractional D3 branes also in the generic orbifold of $\mathbb{C}^2/\Gamma$, and
we will do so in this Section. As we shall see, such a solution clearly encodes
the perturbative behaviour of the world-volume gauge theory in the range in
which perturbation theory makes sense, and presents interesting effects, like
the non-constant flux of the RR 5-form, which is expected to be  related in
some way to the dependence of the degrees of freedom on the scale. 
\par    
The supergravity action we consider has to be a consistent truncation of the
effective action for type IIB strings on $\mathbb{C}^2/\Gamma$ involving only 
the relevant fields for fractional D$3$-branes. These are the metric,  the RR
4-form $C_{4}$, and the twisted fields, as suggested by the linear  couplings
obtained through the boundary  state analysis in Appendix  \ref{accoppia}. The
dilaton and axion fields we take instead to be constant. Our starting point is
thus the bulk action
\beq
\label{bulk1}
S_{\rm b}= \frac{1}{2 \kappa^2} \Bigg\{ \int d^{10} x
\sqrt{-\det G}~R - \frac{1}{2} \int \Big[ G_{3}\wedge {}^* \bar G_{3}
-\frac{\ii}{2}C_4\wedge G_3\wedge \bar G_3  
+\frac{1}{2}{\widetilde{F}}_{5}
\wedge {}^* {\widetilde{F}}_{5}\Big] \Bigg\}~,
\eeq
where we have introduced a complex field notation, defining $\gamma_2\equiv
C_2-\ii B_2$,  $G_{3} \equiv d\gamma_2= F_3-\ii\,H_3$ where $H_3=dB_2$,
$F_3=dC_2$ and $F_5=dC_4$ are respectively the field strengths corresponding to
the NS-NS 2-form potential, and to the 2-form and the 4-form potentials of the
R-R sector.  As usual, the self-duality constraint on the gauge-invariant%
\footnote{The gauge transformations of the RR forms being given compactly by
$\delta C_{p+1} = d\lambda_p + H_3\wedge \lambda_{p-2}$.} field-strength
$\widetilde F_{5}$ defined as  $\widetilde F_{5} = dC_4+C_2\wedge H_3$ has to
be implemented on shell. Finally,  $\kappa=8\,\pi^{7/2}\,g_s\,\alpha'^2 = 2
\pi^{3/2}\, g_s$,  where $g_s$ is the  string coupling constant.  Notice that
in order for the choice of  constant dilaton and axion to be consistent, the
three-form $G_{3}$ must satisfy the condition \cite{Bertolini:2000dk}
\beq
\label{g3}
G_{3}\wedge {}^*G_{3}=0.
\eeq
We search for the solution corresponding to wrapping a D$5$-brane  on a
collection $m^J e_J$ of vanishing two-cycles%
\footnote{For the  D$5$ wrapped on the cycle $e_0=-\sum_i d_i e_i$ the effect
of a background flux $\int_{e_0} \mathcal{F} = 2\pi$ has to be taken into
account.}. In the orbifold limit, this gives indeed a fractional D3 brane to
which open strings attach with  Chan-Paton factors in the representation
$\oplus_J m^J \mathcal{D}_J$.  Consider the harmonic decomposition of the
complex 2-form field,
\beq
\label{ansaga}
\gamma_2=
2\pi\,\gamma_j\, \omega^j~\,\, , \,\,\,{\rm with}\,\,\,\,\,\,\,
\gamma_i\equiv c_i-\ii b_i~,
\eeq     
where the normalizable, anti-self-dual (1,1)-forms $\omega^i$ defined on the ALE
space are  dual to the exceptional cycles $e_i$, in the sense that 
\beq
\label{omegas}
\int_{e_i}\omega^j = \delta^j_i~,\hskip 0.8cm
\int_{\mathcal{M}_\Gamma}\omega^i\wedge\omega^j  = -(C^{-1})^{ij}~.
\eeq 
Substituting the decomposition \eq{ansaga} into \eq{bulk1} one gets the
following expression for the bulk action
\bea
\label{bulk2}
S_{\rm b} & = & \frac{1}{2 \kappa^2} \Bigg\{\int d^{10} x~
\sqrt{-\det G}~ R -\frac{1}{2} \int \Big[  
\frac{1}{4\pi^2}d\gamma_i\wedge {}^* d\bar\gamma_j
\wedge\omega^i\wedge{}^*\omega^j
\nonumber\\
& - &  \frac{\ii}{8\pi^2}C_4\wedge d\gamma_i
\wedge d\bar\gamma_j\wedge\omega^i\wedge\omega^j
\,+\,\frac{1}{2}\, {\widetilde{F}}_{5}
\wedge {}^* {\widetilde{F}}_{5}\Big] \Bigg\}~.
\eea
The scalar fields $b_i=\int_{e_i}B/(2\pi)$ and  $c_i=\int_{e_i}C_2/(2\pi)$ 
couple to the wrapped D5 branes, and correspond thus to appropriate
combinations of the twisted fields in the orbifold limit (see Appendix
\ref{accoppia} for the explicit relations). In fact, let us start with the 
boundary action for a D5-brane with constant axion and dilaton  and wrap it on
the collection of exceptional 2-cycles:
\beq
\label{d5wrapped}
S_{\rm wv} = -\frac{T_5}{\kappa}\left\{ \int_{D3}\int_{m^J e_J}
\sqrt{-\mathrm{det}(G + B )} + \int_{D3}\int_{m^J e_J} 
\left(C_{(6)} + C_{(4)}\wedge B \right)\right\}~,
\eeq
where the tension of a $p$-brane is given by the usual expression $T_p =
\sqrt{\pi}(2\pi\sqrt{\alpha'})^{3-p}$ $=\sqrt{\pi}(2\pi)^{(3-p)/2}$.  The RR
6-form $C_6$ to which the D5 couples directly is the dual of $C_2$, the precise
relation being modified with respect to the na\"\i ve Hodge duality  so as to
be compatible with the equation of motion of $C_2$:
\beq
\label{dual10}
{}^*dC_{2} + C_{4}\wedge H_{3} = -dC_{6}~.
\eeq
The harmonic decomposition 
\beq
\label{c6dec}
C_{6} = 2\pi\,\mathcal{A}_{\,4,j} \wedge \omega^j
\eeq
provides us with four-forms $\mathcal{A}_{\,4,j}$ that are dual in the six flat
dimensions to the scalars $c_i$, namely we have  $d\mathcal{A}_{\,4,j}$=$
{}^{*_6} dc_i - C_{4}\wedge d b_i$~. Carrying out the integrations over the
exceptional cycles, in the limit in which they have vanishing volume, we obtain
the following expression for the world-volume action%
\footnote{We introduce, for notational simplicity,  the non-independent 
twisted field $b_0$  and $\mathcal{A}_{4,0}$ arising by  wrapping on the  cycle
$e_0 = -\sum_i d_i e_i$ (with a non-trivial flux of  $\mathcal{F}$): 
$$ 
b_0 \equiv {1\over 2\pi}\int_{e_0} (B + \mathcal{F}) =   
1 -\sum_j d_j b_j~;\hskip 0.5cm 
\mathcal{A}_{4,0} \equiv {1\over 2\pi} \int_{e_0} C_6 = -\sum_j d_j
\mathcal{A}_{4,j}~. 
$$}:
\beq
\label{boundaryaction}
S_{\rm wv} =  
-\frac{T_3}{\kappa}\left\{ \int_{D3} \sqrt{-\mathrm{det} G }\,
m^J b_J +\int_{D3} C_{(4)}\, m^J b_J +
\int_{D3} m^J \mathcal{A}_{\,4,J}\right\}~,
\eeq 
explicitly exhibiting the coupling to the twisted fields.
\par
From the bulk and boundary actions \eq{bulk1}, \eq{boundaryaction},  one can
straight-forwardly determine the equations of motion for the various fields
emitted by a fractional D$3$-brane. Moreover, we will utilize the standard
black 3-brane \emph{ans\"atz}%
\footnote{The minus sign in the ansatz for $\tilde F_5$ is consistent to  the
sign of the coupling to $C_4$ that we chose in \eq{boundaryaction}, whereas the
plus sign describe the case of fractional anti-branes.} for the untwisted 
fields:
\bea
\label{stan3}
ds^2 & = & H^{-1/2}\eta_{\alpha\beta}dx^\alpha dx^\beta + 
H^{1/2}\delta_{ij}dx^idx^j~,\label{stan1}\\
\widetilde F_5 & = & - d\left(H^{-1}dx^0\wedge...\wedge dx^3\right) - 
{}^*d\left(H^{-1}dx^0\wedge...\wedge dx^3\right)~.\label{stan2}
\eea
\par 
The condition \eq{g3} requires, upon the harmonic decomposition \eq{ansaga},
that $d\gamma_i\wedge {}^{*_6} d\gamma_j\wedge \omega^i\wedge\omega^j=0$.
Introducing the complex coordinate $z\equiv x^4 +\ii x^5$, the above equation
implies that
\beq
\label{ddbar}
\partial_z \gamma_{i} \, \partial_{\bar z}\gamma_{j}(C^{-1})^{ij} = 0~.
\eeq
This constraint is satisfied either choosing $\gamma_i$'s to be analytic or
anti-analytic functions. However consistency requirements impose that
$\gamma_i$  be an analytic function of $z$ for fractional branes having a
positive R-R untwisted charge%
\footnote{Our convention is to define 'positive' the charge of a brane if it
gives a positive R-R untwisted contribution to the $F_5$ flux.},
while it must be anti-analytic in the case of fractional anti-branes, with a
negative R-R untwisted charge. In what follows we will explicitly refer always
to the first case. 
\par
The equations for the twisted fields $\gamma_i$ is%
\footnote{In writing \eq{eq2b} we use the fact that  the entries $\widehat
C_{i0}$ of the extended Cartan matrix are expressed in terms of the
non-extended matrix by $\widehat C_{i0}=-\sum_j C_{ij} d_j$.}
\beq
\label{eq2b}
\partial_a\partial_a \gamma_i +
2\ii T_3\kappa\frac{\widehat C_{iJ} m^J}{4\pi^2}\delta(x^4)\delta(x^5)=0~,
\eeq
with $a=4,5$. Therefore, the twisted fields are just harmonic functions of $z$;
taking into account the explicit values of $T_3$ and $\kappa$, and
including also the non-independent field $\gamma_0$, we have
\beq
\label{twistfields}
\gamma_I(z) = -\ii\, {g_s\over 2\pi}\, \widehat C_{IJ}m^J\,
\ln \frac{z}{\Lambda_I}
\eeq
where $\Lambda_I$ are IR regulators. Remembering \eq{fbs2}, we see that the
twisted fields run logarithmically with the (complexified) scale exactly as
the  coupling constants of the various gauge groups of the world-volume  theory
\cite{Klebanov:2000rd,TAKA}:
\beq
\label{tau1}
\tau_I(z) \equiv \frac{4\pi\ii}{g_{I}^2} + \frac{\theta_{I}}{2\pi} 
= \frac{\ii}{2\pi} \widehat C_{IJ} m^J\, \ln \frac{z}{\Lambda_{I}}
=- {1\over g_s}\,\gamma_I(z) 
\eeq 
having identified the (complexified) energy scale as $z/(2\pi\alpha')=z$.  The
IR regulators correspond in this perspective to the dynamically generated
scales, where the non-perturbative effects in the gauge theory suddenly become
important. On the other hand, the scale $\rho_0$ at which the twisted fields
attain the values \eq{fbs1} appropriate for the orbifold CFT correspond to the
UV cutoff at which the ``bare'' gauge theory, with coupling constants
\beq
\label{classg0}
\bar g_{I}^2 = 4\pi g_s |\Gamma|/d_I~. 
\eeq   
is defined. Indeed, it is in the orbifold CFT that the world-volume theory is
determined to be the $\mathcal{N}=2$ SYM theory we are discussing. From
\eq{tau1} we see that the UV cutoff is related to the dynamically generated
scales by the relation
\beq
\label{fbs4}
\Lambda_I = \rho_0\,\exp\left(-{2\pi\over g_s\, \widehat C_{IJ}m^J} {d_I\over
\Gamma}\right) = \rho_0\,\exp\left( - {8\pi^2\over \bar g_I^2\, 
\widehat C_{IJ}m^J}\right) ~.
\eeq
Let us notice that the agreement at the perturbative level between the twisted
fields and the running gauge coupling constant holds for any configuration
$\{m^I\}$ of fractional branes, with gauge factors U$(m^I)$ UV free, conformal
or IR free according to the values of the coefficients $\widehat C_{IJ}m^J$.
\par
Let us discuss now the behaviour of the untwisted fields.  We restrict our
attention to configuration of branes such that a set of $|\Gamma|-1$
independent twisted fields (let's say, the $\gamma_i$) run to asymptotic
freedom in the UV (or vanish), while the remaining field ($\gamma_0 = -\ii -
\sum_i d_i\gamma_i$) has obviously the opposite running. That is, we require
the positivity of the $\beta$-function coefficients
\beq
\label{ponte1}
\widehat C_{iJ} m^J = C_{ij}(m^j - d^i m^0)= C_{ij}\widetilde m^j~,
\eeq   
where we introduced the notation $\widetilde m^j= m^j - d^i m^0$ which will be
convenient in the sequel. This case generalizes directly the case of the
$\mathbb{C}^2/\mathbb{Z}_2$ orbifold%
\footnote{In this case, with $N+M$ branes of type 1 and $N$ branes of type 0,
the gauge group is $\mathrm{SU}(N+M)\times \mathrm{SU}(N)$. The single
independent twisted field  $\gamma_1(z)= c_1 -\ii b_1$ corresponds to the
running coupling $\tau(z)$ of the UV free $\mathrm{SU}(N+M)$ theory, while the
$\mathrm{SU}(N)$ factor is IR free, as its coupling described by 
$\gamma_0=-\ii-\gamma_1$. } considered in
\cite{Polchinski:2000mx,Aharony,Petrini:2001fk}.
\par
The equation of motion for the RR 5-form involves a boundary term, obtained 
from the world-volume action \eq{boundaryaction}. If one decomposes  the
twisted fields $b_I$ into back--ground value plus fluctuation, tadpole
regularization implies that only the back--ground value  contributes  to the
equation of motion, so the inhomogeneous equation of motions depend on how one
fixes the back-ground value. At the same time, the contributions from the
twisted fields to the five--form  equation of motion  suffer from IR
divergences, which need to be regularized  by means of IR cutoffs. A natural
procedure would be to identify the back-ground value of $b_I$ with  the value
$b_I=d_I/\Gamma$ it assumes in the orbifold background; this value is attained
by the actual solution \eq{twistfields} at the UV cutoff radius $\rho_0$.
Indeed the stringy derivation \cite{Merlatti:2001ne} of the boundary  action
\eq{boundaryaction} is fully justified in the orbifold background only. Our
approach is  to assume that the boundary action nevertheless describes
correctly the relevant degrees of freedom down to the radii $\Lambda_I$ that
correspond to the dynamically generated scales, covering the  entire range in
which the perturbative approach to the world-volume gauge theory makes sense.
\par
Thus, assuming that some sort of enhan\c con mechanism takes place, we imagine
that  the fractional branes of type $i$ dispose themselves on (or near) a
circle  of radius $\Lambda_i$, where their effective tension and charge, 
proportional to $b_i(\Lambda_i)$, vanish, see \eq{boundaryaction} while,  for
the branes of type $0$, generalizing the $\mathbb{Z}_2$ set-up,  we assume them
to be located  in the point in which $\gamma_0(z)=-\ii$.  The tension and the
D3-charge of the branes of type $i$ are however not lost; they are stored in
the  twisted fields. We are assuming that the branes of types $i$ correspond to
UV free gauge theories, so the corresponding dynamically generated scales
$\Lambda_i$ are small. 
\par
With such an interpretation, the equation of motion for the 5-form field
strength takes the form
\beq
\label{eq1}
d{}^* \widetilde F_5= \ii (2\pi^2)\,
d\gamma_i\wedge d\bar\gamma_j (C^{-1})^{ij}\, \wedge\Omega^4
+ 2 T_3\kappa\,m^0\, \Omega^2\wedge\Omega^4~,
\eeq
where the four-form $\Omega_4$ has delta-function support on $x^6=\ldots=x^9=0$
and $\Omega_2$ on $x^4=x^5=0$.  Notice in the r.h.s. we included the  explicit
charge localized on the fractional D3 branes of type $0$ only.  
\par
\par
The (non constant) untwisted charge is measured by the flux  $\Phi_5(\rho)$ of
the RR 5-form $\widetilde F_5$ through a surface which intersects the $z$-plane
in a  circle of radius $\rho$. This can be obtained integrating directly
\eq{eq1} on a region bounded by such a surface. Substituting the twisted fields
\eq{twistfields} we remain with integrals in the $z$-plane which diverge for
small values of the radius. It is natural to regulate them by cutting the
integration at the enhan\c con radii $\Lambda_I$  appropriate for each term in
the r.h.s of \eq{eq1}:  
\bea
\label{fbs3}
\Phi_5(\rho) & = & 
4\pi^2 g_s\left(m^0 + {g_s\over 2\pi}\sum_{i,j} m^M\widehat C_{Mi} 
(C^{-1})^{ij} \widehat C_{jM} m^M \int_{\max (\Lambda_i,\Lambda_j)}^\rho 
{d\rho'\over\rho'}\right)
\nonumber\\
& = & 4\pi^2 g_s\left(m^0 + {g_s\over 2\pi}
\sum_i \theta(\rho-\Lambda_i)\,m^M\widehat C_{Mi} \widetilde m^i\, 
\ln {\rho\over\Lambda_i} \right)~.
\eea
The flux is thus proportional to the untwisted charge $Q = m^I b_I$ encoded in
the boundary action \eq{boundaryaction}. This can be seen, for instance, using
\eq{fbs4} to write  the flux in terms of the  UV cut-off, obtaining, for scales
above the biggest enhan\c con radius, so that we can forget the 
theta-functions in \eq{fbs3},
\beq
\label{flusso} 
\Phi_5(\rho) = 
4\pi^2\,g_s\left(\frac{m^I d_I}{|\Gamma|}
+{g_s\over 2\pi}\widehat C_{IJ}m^I m^J\ln\frac{\rho}{\rho_0}\right)
= 4\pi^2\,g_s\, Q(\rho)~.
\eeq
In this form, the first contribution is attributed to the untwisted D3 charge
localized on the fractional branes, due to the background value  \eq{fbs1}.
This is the picture which most directly compares with the yield of the
couplings to the boundary states described in Appendix \ref{app:b}. The
non-constant (and in particular, logarithmic) RR flux is a  seemingly general
feature of fractional brane solutions in various contexts \cite{KLEBA3}. 
Though it is generally believed to be related to the decrease of degrees of
freedom towards the IR, its precise interpretation seems to depend  on the
specific context. We will discuss some properties of the flux \eq{fbs3} of our
model in the next Section.
\par
Through the  \emph{ans\"atz} Eq.s (\ref{stan1},\ref{stan2})  the R-R untwisted
field equation \eq{eq1} gives rise, consistently with the yield of the Einstein equation
that we  omitted for the sake of shortness, to the following equation for the
function $H$:
\beq
\label{eq1b}
\partial_i\partial_i H +4\pi^2\,\partial_z\gamma_i\partial_{\bar z}\bar\gamma_j
(C^{-1})^{ij}\delta(x^6)...\delta(x^9)
+2T_3\kappa\, m^0\delta(x^4)...\delta(x^9)
=0~,
\eeq
with $i=4,\ldots 9$. The solution of this equation involves IR divergent
integrations on the $z$ plane, which, in perfect analogy to the computation of 
the $\Phi_5$ flux, we regularize by means of the enhan\c con radii.  In the
end, for scales above the highest enhan\c con, the function $H$ can be
expressed in terms of the UV cutoff $\rho_0$ as
\beq
\label{sol2}
H =  1+ {g_s\over \pi}\frac{m^I d_I}{|\Gamma|}
\frac{1}{r^4}+ {g_s^2\over 4\pi^2} \frac{m^I \widehat C_{IJ}m^J}{r^4}
\left[\ln\frac{r^4}{\rho_0^2 \sigma^2}-1+\frac{\rho^2}{\sigma^2}
\right]~,
\eeq
where $\rho^2=(x^4)^2+(x^5)^2$, $\sigma^2 =\sum_{i=6}^9 (x^i)^2$  and
$r^2=\rho^2+\sigma^2$. The first contribution would arise by a source term for
$C_4$ in \eq{eq1}, localized on the fractional branes and due to the orbifold
background values  \eq{fbs1} of the twisted fields. 
\par
The curvature scalar for the metric of \eq{stan1} takes the following simple
form 
\begin{equation}
\label{curv}
R =
-{1\over
2}H^{-3/2}\partial_i\partial_i H
\end{equation} 
and hence vanishes outside the fixed plane where the different sources for the
function $H$ live (see \eq{eq1b}). The metric exhibits  singularities whenever
$H$ vanishes and an horizon on the fixed plane. In Appendix \ref{app:a} we will
briefly discuss further the form of $H$. 
\section{Relation to the gauge theory}
\begin{figure}
\null
\begin{center}
\epsfig{file=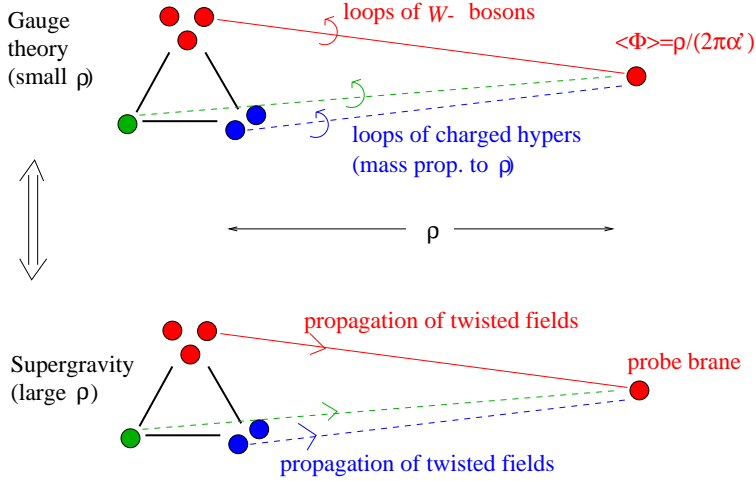,width=10cm}
\caption{The dual interpretation of a fractional brane configuration
corresponding, on the one hand, to a particular corner of the Coulomb phase 
moduli space in the gauge theory and, on the other hand,  to a single brane
probe in the classical supergravity background} 
\label{fig2}
\end{center}
\end{figure}

We already put forward in \eq{tau1} the main relation between the classical
supergravity solution and the quantum behaviour of the world-volume gauge
theory, namely the identification between the twisted scalars and the one-loop
running  coupling constants. This relation can be understood, and attributed
substantially to the usual open-closed string duality of the ``cylinder''
diagrams, by using a probe-brane approach.  Let us thus consider the set-up
described in Figure~\ref{fig2}, in which a single test fractional brane of type
$I$ moves slowly in the supergravity  background generated by a configuration
of $m_J$ branes in each representation $\mathcal{R}_J$ of a given orbifold
group $\Gamma$.
\par
Let us expand the world-volume action \eq{boundaryaction}  for our test brane 
up to quadratic order in the 'velocities' $\frac{\partial
x^a}{\partial\xi^\alpha}$, where $a=4,5$ are the transverse directions
accessible to our fractional probe and $\xi^\alpha$ are the world-volume
coordinates. The dependence from the function $H$ cancels at this order, and
the part of order zero in the velocities is a constant, so that no potential 
opposes the motion of our probe in the $x^4,x^5$ directions, \emph{i.e.}, in
the $z$-plane. In the dual gauge theory picture, this BPS property corresponds
to the opening up of the Coulomb branch of the moduli  space.  As usual, the
kinetic terms of the probe action,
\beq
\label{probeact1}
S_{{\rm p},I} \sim 
{T_3\over 2\kappa} \int d^4\xi\, 
\left(\frac{\partial x^a}{\partial\xi^\alpha}\right)^2\, 
b_I~,
\eeq
should describe the  effective metric on this moduli space.  
\par
On the gauge theory side, the moduli space for the $U(m^I+1)$ gauge factor
associated to the branes of type $I$ is   parametrized by the expectation
values of the adjoint complex scalar in the  vector multiplet,
$\langle\Phi\rangle =  \mathrm{diag}(a_1,\ldots,a_{m_I+1})$,   which 
generically break the gauge group to U$(1)^{m_I+1}$,  and by the masses $M_k$
($k=1,\ldots N_{\rm h}$, with $N_{\rm h} = \widehat A_{IJ} m^J$) of the
fundamental hyper-multiplets. As depicted in Figure \ref{fig2}, our probe brane
configuration  corresponds to the corner of the moduli space where one of the
$a$'s,  say $a_{m_I+1}$, assumes a large value $z$, with $|z| \gg
\Lambda_{(I)}$. Moreover  the (complexified) masses $M_k$ of the hyper-multiplet
fields, which are the lowest excitations of the open strings stretching from
the probe to other types of fractional branes sitting at the origin, also are
all equal to $z$. 
\par
Substituting in \eq{probeact1} the classical solution \eq{twistfields} for the
twisted field $b_I$  and identifying $z\equiv  x^4 +\ii x^5$ with the v.e.v
$a_{m_I+1}$ of the Higgs field, as required by the open string description of
the gauge theory living on the branes, we obtain the kinetic term 
\beq
\label{sproberun}
S_{{\rm p},I} = 
\int d^4\xi\, \frac{\widehat C_{IJ}m^J}{8\pi^2}
\ln \frac{\rho}{\Lambda_I}\,\, 
\left|\partial_\alpha{a_{m_I+1}}\right|^2~.
\eeq
We see that  the effective tension of the probe brane moving in the $z$ plane,
namely the effective metric on the moduli space corner parametrized by 
$a_{m_I+1}$, coincides with the running coupling constant $1/g_I^2(\rho)$ of
\eq{tau1} for the $U(m^I)$ gauge theory of the ``background'' branes of type
$I$. If, instead of considering only the terms in \eq{boundaryaction}, we 
expand the full world-volume action for the probe brane, keeping track also of
the world-volume gauge field $\mathcal{F}$, we find consistently that the
resulting gauge coupling and theta-angle for the world-volume U$(1)$ field are
summarized in the $\tau_I(z)$ of \eq{tau1}.  
\par
This is also the correct result from the field theory side. Indeed,  the
perturbative%
\footnote{There are no perturbative corrections from higher loops, but only 
instanton corrections.} part of the  pre-potential  of the $\mathcal{N}=2$
low-energy effective theory, which contains one-loop effects,  is given by
(see, \emph{e.g.}, \cite{D'Hoker})
\bea
\label{prepot}
\mathcal{F}_{\rm pert} &=& 
-\frac{1}{8\pi\ii}\left(\sum_{p\not= q=1}^{m_I+ 1}
(a_p - a_q)^2\ln \frac{(a_p - a_q)^2}{\Lambda_{I}^2}\right. 
\nonumber\\
& &\phantom{-\frac{1}{8\pi\ii}\Bigl(}\left. 
- \sum_{p=1}^{m_I+ 1}\sum_{k=1}^{N_{\rm h}} 
(a_p + M_k)^2\ln \frac{(a_p +M_k)^2}{\Lambda_{I}^2}\right)~.  
\eea
The effect of the non-trivial vev $a_{m_{I+1}}$, corresponding to the position 
of the probe brane, is to Higgs  U$(m^I+1)$ into U$(m^I)\times\mathrm{U}(1)$.
The effective coupling of the U$(1)$ (which also appears in the effective 
action as moduli space metric) is given by
\beq
\label{fbs5}
\frac{\partial^2\mathcal{F}}{\partial a_{m_{I+1}}^2}\sim 
\frac{\ii}{2\pi} (2 m^I - N_{\rm h})\,\ln \frac{z}{\Lambda_{I}}~,
\eeq
namely it coincides indeed with $\tau_I(z)$. 
\par
The origin of the identification \eq{tau1} between the twisted scalar
$\gamma_I$ and the effective coupling $\tau_I$ is quite simply understood by
considering the one-loop diagrams that contribute to the  perturbative
pre-potential \eq{prepot}, as summarized in Figure \ref{fig2}. Each such diagram
arise in the field-theory limit from the one-loop amplitude of open strings
attached on one side to the probe brane and on the other side to a background
brane of the same type (then $W$-bosons multiplets  run in  the loop) or of
other types (then hyper-multiplets run in the loop). Upon open-closed duality,
these cylinder diagrams are seen as tree-level closed string exchange diagrams
which, in the supergravity limit, describe the interaction of the probe brane
with the twisted fields emitted by the background.   
\par
To go beyond the perturbative level (which task we will not attempt here),  on
the field-theory side one may construct the Seiberg-Witten curve,  as it is
done in \cite{Petrini:2001fk} in the $\mathbb{C}^2/\mathbb{Z}_2$ case.  In the
moduli space corner corresponding to our probe brane configuration, the
effective theory will receive corrections proportional to powers  of the
one-instanton contribution to the partition function 
\beq
\label{dp1}
\exp\left(-{8\pi^2\over g^2_I} +\ii \theta_I\right) = \exp
\left(2\pi\ii\tau_I\right) =
\left({\Lambda_I\over z}\right)^{\widehat C_{IJ}m^J}~.
\eeq
On the string theory side, such effects are likely due \cite{Klebanov:2000rd}
to fractional D-instantons, namely to D1 (Euclidean) branes wrapped on the
vanishing  cycles $e_i$, with action 
\beq
\label{dp2}
\exp \left(-{2\pi\ii\over g_s}\gamma_I\right) = \exp
\left(2\pi\ii\tau_I\right)~.
\eeq
It would be very interesting to pursue  this argument further and determine the
coefficients of the instanton corrections.
\par
The instanton corrections in the field theory description appear to be
consistent with the choice of using the dynamically generated scales
$\Lambda_i$ as IR regulators for the contributions of the twisted fields to the
untwisted supergravity equations of motion. Indeed, especially for large
numbers of  fractional branes, as pointed out in 
\cite{Polchinski:2000mx,Petrini:2001fk},  the instantonic contributions 
\eq{dp1} become quite suddenly important near  the enhan\c con radius
$|z|=\Lambda_i$ and the form of the Seiberg-Witten curve suggests that for
smaller radii the coupling $\tau_i$ stops running. Then \eq{tau1} would be
valid for $|z|>\Lambda_i$ only, while for $|z|<\Lambda_i$ $\gamma_i$ would be
constant so that $\partial \gamma_i =0$. This agrees with the cut off at 
$\rho\equiv|z|=\Lambda_i$ of the twisted contributions (always proportional to
derivatives of the $\gamma$'s) to the RR five-form flux,  \eq{fbs3}, and to
$H$, \eq{sol2}.   
\par
What field theory interpretation can be given of the behaviour of the untwisted
fields? Namely, what information can be deduced from the full supergravity
solution, and from its embedding in a consistent string theory?  As mentioned
in the introduction, interesting proposals and discussions, all focused on the
$\mathbb{C}^2/\mathbb{Z}_2$ case, have appeared in the literature
\cite{Polchinski:2000mx,Aharony,Petrini:2001fk}.   In the
$\mathbb{C}^2/\mathbb{Z}_2$ case, at the enhan\c con radius $\Lambda$ the
$b$ field, and so the untwisted charge of each brane of type 1, changes sign:
below the enhan\c con, the system is no longer BPS. A BPS situation can be
recovered by shifting  $b\to b + 1$ which, as discussed after \eq{fbs1}, 
corresponds to a symmetry of the string background. Nevertheless, to preserve
the overall untwisted charge $Q$,  $M$ bulk branes must be subtracted,
modifying the gauge group to  $\mathrm{SU}(N)\times \mathrm{SU}(N-M)$. 
\par
In \cite{Polchinski:2000mx} it is suggested that this picture implies a
duality akin to the one discovered in \cite{KLEBA3} in the context of the
$\mathcal{N}=1$ theories  associated to fractional branes at the conifold.
Below the enhan\c con scale,  the theory is dual to a theory with $\tau' = \tau
+ \ii/g_s$, reduced gauge groups and a dynamically generated scale $\Lambda'$
rescaled by a factor of $\exp(-\pi/g_s M)$ with respect to $\Lambda$. Below
$\Lambda'$ the duality is to a theory with further reduced gauge groups
(corresponding again to the removal of $M$ bulk branes), and so forth, until
one cannot subtract any more $M$ bulk branes. 
\par
The change of the gauge group between scales related by a factor of
$\exp(-\pi/g_s M)$ is instead interpreted in \cite{Aharony} as due to a 
Higgs phenomenon implying that  $M$ bulk branes are actually distributed 
within such two scales.
\par
In \cite{Petrini:2001fk} a more conservative point of view is supported; based
on the comparison with the appropriate Seiberg-Witten curve which they construct
explicitly, these authors argue that it is not necessary to try to extend the
supergravity description below the ``first'' enhan\c con $\Lambda_1$, where
(D--)instanton effects become all important both in field theory and in string 
theory.     
\begin{figure}
\begin{minipage}[b]{0.48\linewidth}
\centering \includegraphics[width=6.7cm]{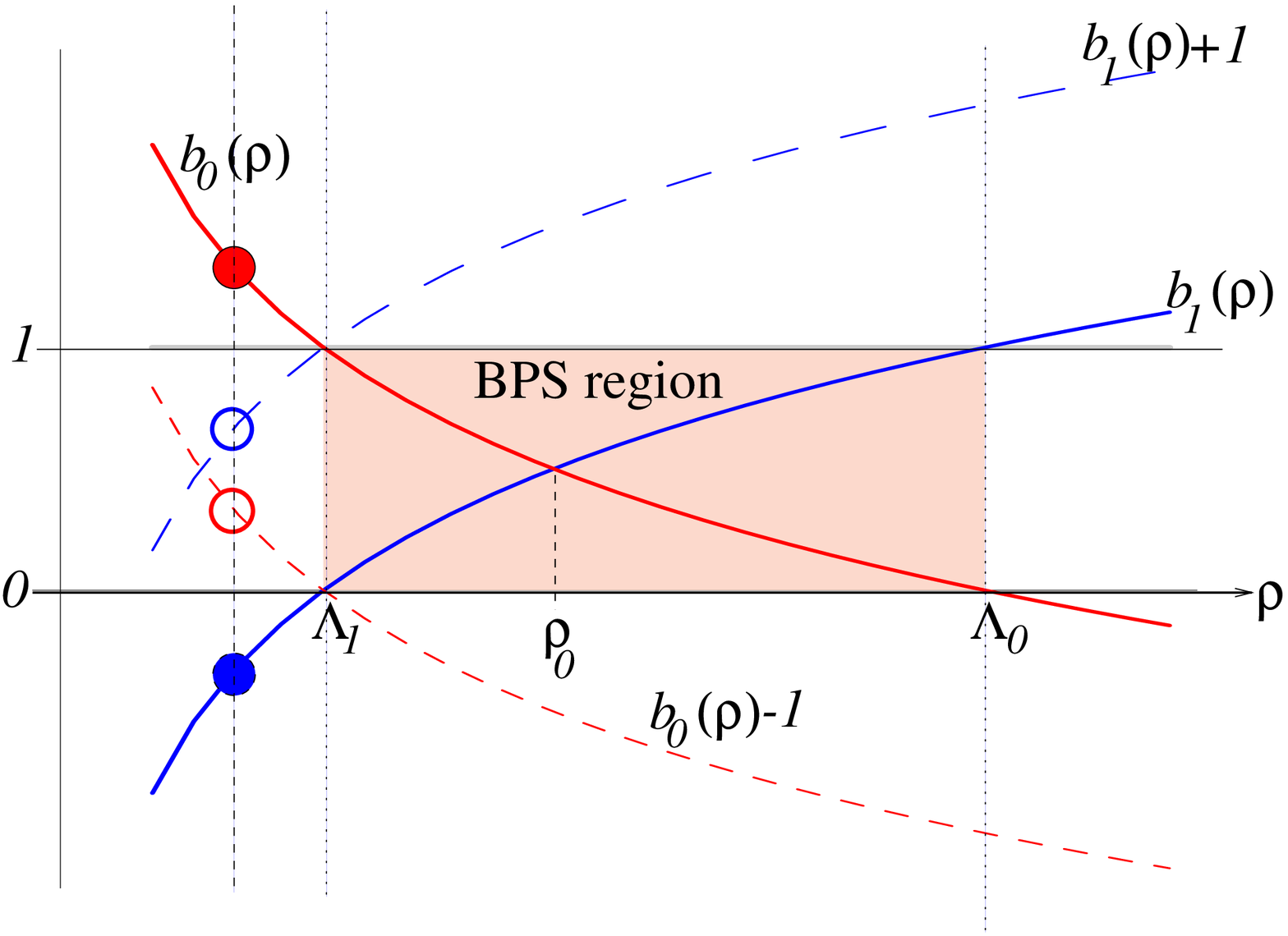}
\caption{For the $\mathbb{C}^2/\mathbb{Z}_2$ orbifold, below the enhan\c con
$\Lambda_1$, where the fields $b_I$ assume non-BPS values 
(indicated by a filled circle) one can, by unit shifts to the values indicated by an
empty circle, reach a BPS regime.}\label{fig:bz2}
\end{minipage}
\hskip 0.2cm
\begin{minipage}[b]{0.48\linewidth}
\centering \includegraphics[width=6.7cm]{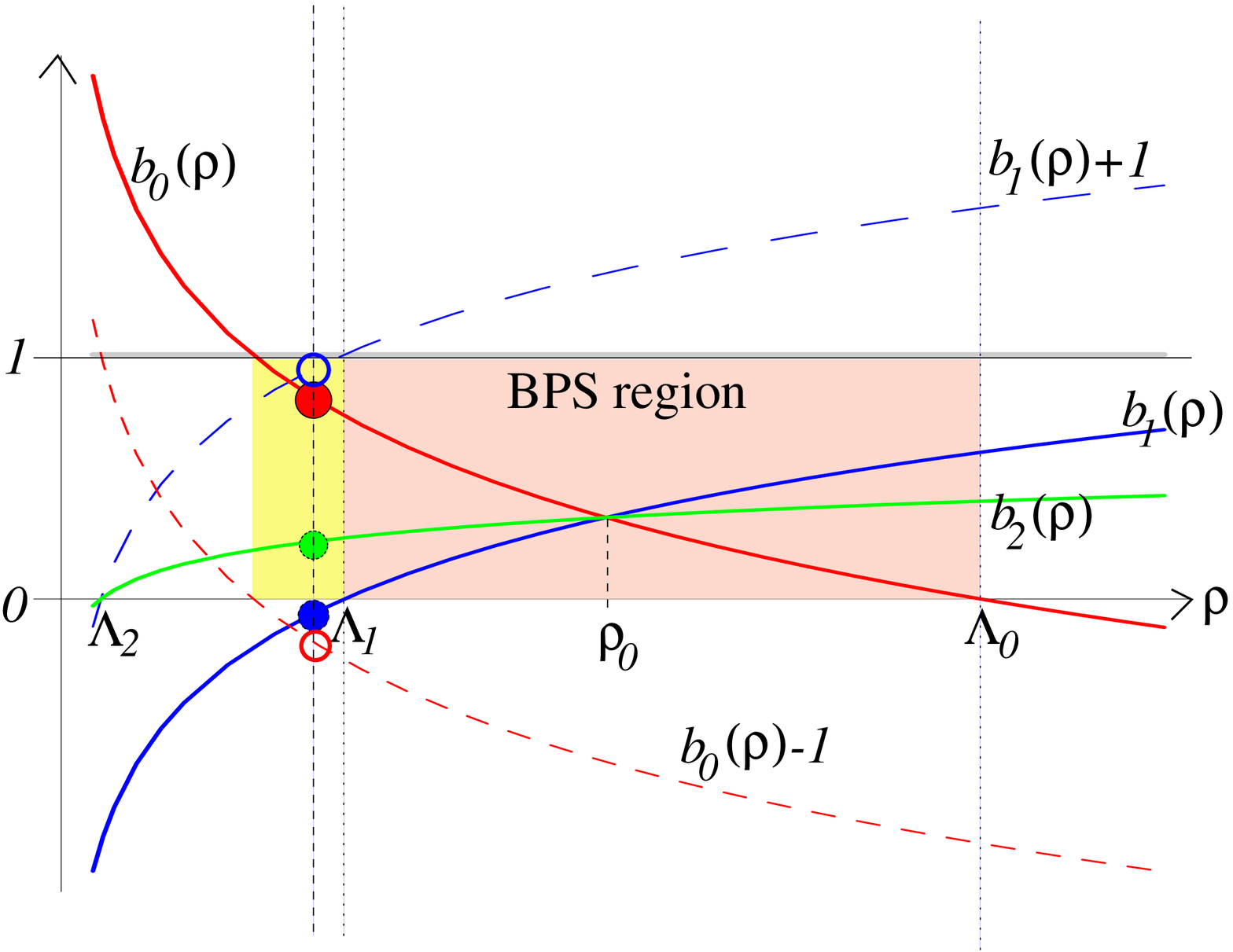}
\caption{For the $\mathbb{C}^2/\mathbb{Z}_3$ orbifold, immediately below the
first enhan\c con $\Lambda_1$, the non-BPS values (filled circles) cannot be 
turned by unit shifts (leading, e.g., to the values indicated by the empty
circles) all into positive values.}\label{fig:bz3}
\end{minipage}
\end{figure}
\par
What changes for a generic orbifold $\mathbb{C}^2/\Gamma$?  Having in general
more than two gauge factors, it turns out that when a $b_i$ field becomes
negative (namely, at a radius below its enhan\c con $\lambda_i$), making the
system non BPS, it is not always possible  to recover a BPS system by shifting
other $b_j$-fields by integers. As an example,  consider the 
$\mathbb{C}^2/\mathbb{Z}_3$ orbifold with three gauge groups,  two of them (say
1 and 2) being UV free while the last one (say 0) is IR free%
\footnote{Such a situation may be realized with appropriate choices of the 
$m^I$, for example $(m^0=1,m^1=4,m^2=3)$.}. In such a situation, one has  (see
figure) 
\begin{equation}
\label{run1}
b_0+b_1+b_2=1~, \,\, \Lambda_2<\Lambda_1<\Lambda_0~.
\end{equation}
In the region where $\rho \lesssim\Lambda_1$, $b_1$ is small,  negative while
$0<b_0,b_2<1$ and the system is not BPS. The shift  $b_1 \to b_1+1$ has to be
compensated with a negative shift of $b_0$ (or  $b_2$). Such a shift would
however render this latter field negative, so that  one goes from a non-BPS
situation to another non-BPS situation.  It appears therefore that the
possibility of relating the theory below an enhan\c con to another theory via a
duality based on the periodicity of the $B$-fluxes \cite{Polchinski:2000mx}
does not apply to the generic situation. 
\par
Let us notice that the 5-form flux $\Phi_5(\rho)$ is always  decreasing
towards the IR with the scale $\rho$, for any configuration of fractional
branes. In fact, the coefficient of the logarithm in \eq{flusso} satisfies
\beq
\label{mercul1}
\widehat C_{IJ} m^I m^J\geq 0
\eeq
for any set of $m^I$'s, since the extended Cartan matrices of the ADE series
are positive semi--definite. Let us moreover notice that the flux \eq{flusso}
or, equivalently, the  charge $Q(\rho)$ satisfies the differential equation
\beq
\label{mercul2}
{d Q\over d\ln\rho} = {g_s\over 2\pi}\,\widehat C_{IJ} m^I m^J
\propto \,\beta_i  \bar\beta_j\,G^{ij}~,
\eeq
where we introduced the logarithmic derivatives
\beq
\label{mercul4}
\beta_i \equiv{d\tau_i\over d\ln\rho} = -{1\over g_s} 
{d\gamma_i\over d\ln\rho} =
{\ii\over 2\pi}\,\widehat C_{iJ} m^J 
\eeq
of the ``twisted'' supergravity scalars, which represent the beta-functions 
for the gauge theory couplings $\tau_i$ dual to these supergravity fields%
\footnote{Recall the definition \eq{tau1} of the couplings $\tau_i$. Then 
${\rm Im}\beta_i = 4\pi (-1/ g^3_i)(d g_i/d\ln\rho)$ is related to the usual
beta-function $d g_i/d\ln\rho= - b_1^{(i)}/g^3_i$ by  ${\rm Im}\beta_i
=b_1^{(i)}/ 2\pi$. Using \eq{fbs2}, this agrees with \eq{mercul4}.} and we
denoted by $G^{ij}$ the  metric appearing in their kinetic term. Indeed,
carrying out the integration over the ALE space of the bulk action \eq{bulk2},
one finds a 6-dimensional kinetic term proportional to 
\beq
\label{mercul5}
\int d^6x\, \sqrt{-g_6}\, \partial_\mu \gamma_i
\partial^\mu \bar\gamma_j\,(C^{-1})^{ij}~,
\eeq   
identifying the kinetic matrix as $G^{ij}\propto (C^{-1})^{ij}$. The equation
\eq{mercul2} satisfied by the untwisted charge $Q(\rho)$ is reminiscent of the
equation for the holographic $c$-function of \cite{Anselmi:2000fu}, which with
the same notations (in a different context!) reads
\beq
\label{mercul6}
{d {\ln c}\over d\ln\rho} = 2\beta_i \beta_j\, G^{ij}~. 
\eeq 
Of course, this equation arises in the context of 5D supergravity, and the
logarithmic derivative in the l.h.s. is justified only upon the identification 
between the scale factor of the 5D metric and $\ln\rho$, which holds at
criticality only \cite{Anselmi:2000fu}. The similarity with equation
\eq{mercul2} is thus mainly formal; nevertheless, it would be interesting to
understand whether the relation of the charge $Q(\rho)$, i.e., of the 5-form
flux, with the logarithm of a (holographic?) $c$-function could be more
substantial. 
\subsection*{Acknowledgments} 
We thank M. Bertolini, P. Fr\'e, M.L. Frau, A. Lerda, R. Musto, I. Pesando and R. Russo 
for enlightening discussions. L.G. thanks the Laboratoire de 
physique de L'ENS-Lyon (France) for hospitality, A.L. thanks
Universit\`a degli Studi di Napoli for the same reason.
\appendix
\section{Analytic study of the function $H$}
\label{app:a}
We shall give here a brief analytical study of the function $H$  characterizing
the ansatz Eq.s (\ref{stan1},\ref{stan2}), whose expression  was given in
\eq{sol2}.  We shall draw, in part qualitatively, its level curves. With
reference to the notations of  Section 3, let us define the  two positive real
variables 
\begin{equation}
x=\rho^2 > 0 , \,\, y= \sigma^2 >0~.
\end{equation}
We rewrite then for commodity the function $H$ as follows:
\begin{equation}
H(x,y) = 1+{A\over (x+y)^2} +
{B\over (x+y)^2}\left( \mathrm{ln}\left({(x+y)^2\over \rho_0^2 y}\right)
-1+{x \over y}  \right)~,
\end{equation}
where $A$ and $B$ are two strictly positive constants, which can be deduced by
comparison with \eq{sol2}. The partial 
derivatives with respect to $x$ and $y$ are given by
\beq
\partial_x H = -{2(H-1)\over x+y} +B {3y+x\over y(x+y)^3}\,\,\,\,\,;\,\,\,\,\,
\partial_y H = -{2(H-1)\over x+y} +B {2y^2-(x+y)^2\over y^2(x+y)^3}~.
\eeq 
These expressions imply that  there is neither a local
extremum, nor a multiple point for $H$ in the open quarter plane. Such a 
point, if it exists, has to be located on the axes $x=0$ or $y=0$. Indeed, 
such a point is characterized by a vanishing gradient, which in turn requires 
that
\begin{equation}
(x+y)(x+2y)=0~,
\end{equation}
a condition impossible to satisfy for $x$, $y>0$.
\paragraph{H on the axis $x=0$}
From the previous expressions, we deduce that the function $H(0,y)$
increases monotonically from $-\infty$ in $y=0$ to a maximum $\chi>1$ reached 
in $y_{\chi}$. Let us also denote by $y_1$ the point on the $x$ axis such that
$H(0,y_1)=1$. Explicitly, one finds
\begin{equation}
y_1 = \rho_0^2 \mathrm{exp}\left( 1-{A\over B}\right) \, ,\, 
y_{\chi}=\sqrt{e}y_1 \, , \,
\chi=1 +{B\over 2 \rho_0^4}\mathrm{exp}\left( {2A\over B}-3\right)~.
\end{equation} 
For $y>y_{\chi}$, $H(0,y)$ decreases monotonically and one has $\lim_{y\rightarrow
+\infty} H(0,y)=1$. 
\paragraph{H at constant $x>0$}
In that case one has 
\begin{equation}
\lim_{y\rightarrow 0} H(x,y)=+\infty \,\,\,\, \mathrm{ and} \,\,\,\, 
\lim_{y\rightarrow +\infty} H(x,y)=1~.
\end{equation} 
On the one hand, this tells us that 
$H$ has no definite limit in $(0,0)$. On the other hand, one would like to
know if $H$ is single-valued on this axis. To this purpose, let us introduce the
auxiliary function
\begin{equation}
Q(x,y)= - { (x+y)^3\over 2}\partial_y H = A + B \left[ \mathrm{ln}\left(
{(x+y)^2\over \rho_0^2 y}\right)+ {x^2 +4xy-3y^2\over 2y^2}\right] 
\end{equation}
so that the question translates in the knowledge of the sign of $Q$ at $x$
fixed. The $y$-derivative of $Q$ reads
\begin{equation}
\partial_y Q = {B\over (x+y)y^3}\left( 2y^3 - (x+y)^3\right)~.
\end{equation}
From this expression, it follows that $Q$
decreases from $+\infty$ in $y=0$ to a minimum 
in  $y ={x\over 2^{1/3} -1}$.
Above this point, $Q$ increases so that $\lim_{y\rightarrow
+\infty} Q(x,y)=+\infty$. Direct inspection shows that for 
\begin{equation}
X= y_1 {2^{1/3} -1\over 2^{2/3}} \mathrm{exp}\left( -(
2^{1/6}-2^{-1/6})^2\right) \, , \, Y={X\over 2^{1/3} -1}
\end{equation}
one has 
\begin{equation}
Q(X, Y) = 0 \,\,\,\,\,{\rm and}\,\,\,\,
Q(x, {x\over 2^{1/3} -1}) \gtrless 0  
\,\,\,\,{\rm for}\,\,\,\,\, x \gtrless X~.
\end{equation}
Hence, for $x \geq X$, $H(x,y)$ monotonically decreases while for 
$x < X$, it reaches a local minimum and then a local maximum and so 
is not single valued. One has 
\begin{equation}
\mu=H(X,Y) = 1 + (\chi -1 )2^{4/3}(2^{1/3} -1)
\mathrm{exp}\left( (
2^{1/6}-2^{-1/6})^2-1\right) \, , \, \chi>\mu>1.
\end{equation}
\begin{figure}
\null
\begin{center}
\epsfig{file=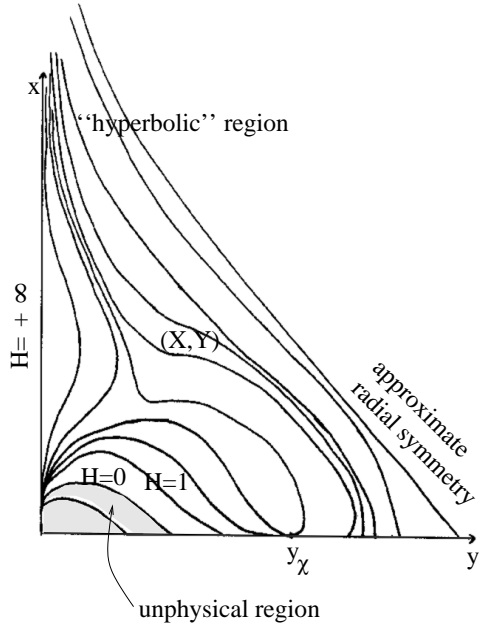,width=6.5cm}
\caption{Level curves of the function $H(x,y)$. See the text for the meaning of
the special points indicated in the figure.}
\label{fig4}
\end{center}
\end{figure}
\paragraph{Level curves}
From the previous observations, we can reach a qualitative understanding 
of the level curves $H(x,y)=h$ which separate in three families as shown on 
Figure \ref{fig4}. 
To the first family belong the curves with $\chi>h>-\infty$ stretching from the
origin point $(0,0)$ and the axis $x=0$ for $-\infty<y<y_{\chi}$. To the second
belong those with $h\geq\chi$, starting at $(0,0)$ and going in the region
where $x \gg X$. The  last family contains the level curves  starting from the
axis  $x=0$ at a position $y>y_{\chi}$ (with $\chi > h> 1$) and going in the
region  where $x \gg X$.
Near the origin point $(0,0)$, any level curve $H(x,y)=h$ may be 
represented by 
\begin{equation}
x=y F_{h}(y) \, , \, F_{h}(y) \approx |\mathrm{ln}y|.
\end{equation}
In the region where $x \gg X$, the level curves are asymptotically 
hyperboles of equation 
\begin{equation}
xy = {B\over h-1}.
\end{equation}
In the region where $x \gg X$ and $y \gg y_{\chi}$, the level curves are 
asymptotically straight lines of equation
\begin{equation} 
x+y = y_{h} \, , \, H(0,y_h)=h.
\end{equation} 
In this last region,we have an approximate six-dimensional spherical 
symmetry.
\section{Closed strings on $\mathbb{C}^2/\Gamma$}
\label{app:b}
\subsection{Bulk massless spectrum in the orbifold theory}
The closed string theory on the orbifold space $\mathbb{C}^2/\Gamma$ admits
untwisted and twisted sectors. Let us consider (the conjugacy class of) an
element $g\in\Gamma$ of order $N$. We have then $N$ 
twisted sectors $T_k$, with $k \in [0,N-1[$; $k=0$ labels the untwisted sector.
We define $\nu = k/N \in [0,1[$. The conformal theory is as usual along the directions 
$0,1,\cdots,5$ while it depends on the twist in the orbifolded directions. 
In a given sector, the modings of the world-sheet complex fields $X^l$ and
$\psi^l$ ($l=1,2$) are given in Table \ref{modings}.
\begin{table}
  \begin{center}
\begin{tabular}{ccc|ccc|ccc}
\hline\hline
\multicolumn{3}{c}{Bosons} & \multicolumn{3}{c}{R-R fermions}
& \multicolumn{3}{c}{NS-NS fermions} \cr   
\hline
\null & $l=1$ & $l=2$ & \null & $l=1$ & $l=2$ &  \null & $l=1$ & $l=2$\cr
\hline\hline
$X^l$ & $\Z+\nu$ & $\Z-\nu$ & $\psi^l$ & $\Z+\nu$ & $\Z-\nu$ &
$\psi^l$ & $\Z+{1\over 2}+\nu$ & $\Z+{1\over 2}-\nu$ \cr
$\bar X^l$ & $\Z-\nu$ & $\Z+\nu$ & $\bar\psi^l$ & $\Z-\nu$ & $\Z+\nu$ &
$\bar\psi^l$ & $\Z+{1\over 2}-\nu$ & $\Z+{1\over 2}+\nu$ \cr
$\tilde X^l$ & $\Z-\nu$ & $\Z+\nu$ & $\tilde\psi^l$ & $\Z-\nu$ & $\Z+\nu$ &
$\tilde\psi^l$ & $\Z+{1\over 2}-\nu$ & $\Z+{1\over 2}+\nu$ \cr
$\tilde{\bar X}^l$ & $\Z+\nu$ & $\Z-\nu$ & $\tilde{\bar\psi}^l$ & $\Z+\nu$ & 
$\Z-\nu$ & $\tilde{\bar\psi}^l$ & $\Z+{1\over 2}+\nu$ & $\Z+{1\over 2}-\nu$
\cr 
\hline\hline
\end{tabular}
\caption{Modings of the closed superstring fields in the sector twisted by 
$\nu$.}
\label{modings}
\end{center}
\end{table}
The intercepts may be evaluated directly by computing the Virasoro algebra 
from the basic commutation relations between oscillators.
\par
Let us recall that  he  $SO(4)\sim \mathrm{SU}(2)_-
\otimes \mathrm{SU}(2)_+$ symmetry of the covering space $\mathbb{C}^2$, acting 
on the complex coordinates by
\beq
\label{SO4action}
\left(\matrix{ z^1 & \ii \bar z^2\cr \ii z^2 & \bar z^1}\right) 
\mapsto \mathcal{U}_- 
\left(\matrix{ z^1 & \ii \bar z^2\cr\ii z^2 & \bar z^1}\right) 
\mathcal{U}_+ ~,
\eeq   
with $\mathcal{U}_\mp\in \mathrm{SU}(2)_\mp$, is broken to SU$(2)_+$ by the
action of $\Gamma$ in \eqn{Gammaaction}. The spectrum, and in particular the
massless spectrum, will be organized in representation of the residual symmetry
group SU$(2)_+$. 
\paragraph{The untwisted sector}
The untwisted sector contains the states of the IIB theory that are invariant 
under the orbifold. Such states have a 10-dimensional dynamics.
In particular in the NS-NS sector, the massless states 
\begin{equation}
\psi^{\mu}_{-{1\over 2}}\tilde\psi^{\nu}_{-{1\over 2}}|0,0\rangle~,
\end{equation}
with $\mu,\nu=0,\ldots 5$, are invariant. They correspond, from the 6-dimensional point of
view, to a dilaton, a metric and a Kalb-Ramond field $B_{\mu\nu}$, and they
obviously are singlets of the ``internal'' residual symmetry group SU$(2)_{+}$.
Also invariant are the following 16 6-dimensional scalars 
\begin{equation}
\psi^{a}_{-{1\over 2}}\tilde{\psi}^{b}_{-{1\over 2}}|0,0\rangle~, 
\end{equation}
with $a,b=6,7,8,9$. These states are in the 
$\mathbf{4}\otimes\mathbf{4}$ of the SO$(4)$ symmetry group of the covering 
space, that is to say in the
$[(\mathbf{2},\mathbf{1})\oplus(\mathbf{1},\mathbf{2})]
\otimes
[(\mathbf{2},\mathbf{1})\oplus(\mathbf{1},\mathbf{2})]$
of $\mathrm{SU}(2)_{-}\otimes \mathrm{SU}(2)_{+}$. 
In terms of the residual $\mathrm{SU}(2)_{+}$, 
these states thus organize as follows 
\begin{equation}
\mathbf{1}^{\otimes 5}\oplus \mathbf{2}^{\otimes 4} \oplus \mathbf{3}.
\end{equation}
The total number of invariant massless states in the NS sector is thus 
1+9+6+16=32.
\par
The massless R-R sector of the IIB theory contains a zero-form potential 
$C_{(0)}$, an antisymmetric two form $C_{(2)}$ and an antisymmetric self-dual 
four form $C_{(4)}$. From the 6-dimensional point of view, 
$C_{(0)}$ is invariant and thus gives a 6-d zero-form in the $\mathbf{1}$ of 
$SU(2)_{+}$. $C_{(2)}$ gives a 6-d two form in the 
$\mathbf{1}$ of $\mathrm{SU}(2)_{+}$ and 6 6-d zero-forms organizing in the 
adjoint of the broken $\mathrm{SO}(4)$, hence in the 
$\mathbf{1}^{\otimes 3}\oplus \mathbf{3}$
of the residual $\mathrm{SU}(2)_{+}$. Finally, $C_{(4)}$ 
gives a 6-d four-form and a 6-d zero-form, both in the $\mathbf{1}$ and 
related by self-duality, and three 6-d 
2-forms organizing in the $\mathbf{3}$. This gives a total of 
$(1)+(6+6)+(1+3 \times 6)=32$ states.
\paragraph{The NS-NS twisted sectors}
There are no bosonic zero modes in the orbifolded directions, so the twisted
fields have a six dimensional dynamics. The intercept being
$a_{NS}= |1/2-\nu|$, three cases have to be considered, depending on the value of $\nu$.
\par
In orbifolds of even order, we may consider sectors twisted by an element of
order 2, i.e., we may have $\nu=1/2$. The intercept vanishes and the
world-sheet fermions have zero modes in the orbifolded directions. The 
massless fundamental state is then a bi-spinor of SO$(4)$. On each side, the GSO 
projection selects a spinor of positive chirality and we end with a 
space-time scalar in the $\mathbf{1}$ of $SU(2)_{+}$ and three space-time 
scalars in the $\mathbf{3}$ of $\mathrm{SU}(2)_{+}$.
\par
When we have $0< \nu < 1/2$,
the intercept is positive and there are no fermionic zero modes. Hence the 
non degenerated fundamental state is tachyonic and odd under the GSO projection.
There exist massless states obtained by the action of some complex fermion 
modes on the fundamental one that we shall write in a
vector notation.
Considering the action of $\mathrm{SU}(2)_{+}$ in (\ref{Gammaaction}), 
one obtains the two following $\mathrm{SU}(2)_{+}$ doublets
${\bar z^1\choose -\ii  z^2}$ and ${-\ii \bar z^2\choose z^1}$.
Extending this property from world-sheet bosons to world-sheet fermions, we 
obtain the following four doublets
\beq
\label{fdoublets}
\Psi^{<} = {1\over \sqrt{2}} { \bar \psi^1 \choose -\ii  \psi^2}~,~~
\Psi^{>} ={1\over \sqrt{2}}{ -\ii \bar \psi^2 \choose  \psi^1}~,~~
\tilde{\Psi}^{<} = 
{1\over \sqrt{2}}{ \tilde{\bar \psi}^1 \choose -\ii  \tilde{\psi}^2}~,~~
\tilde{\Psi}^{>} =
{1\over \sqrt{2}}{ -\ii \tilde{\bar \psi}^2 \choose  \tilde{\psi}^1}~,
\eeq
where in each doublet, the two complex fermions have the same modings. 
The four massless states may be thus written in the matrix form
\begin{equation}
\Psi^{>}_{-{1\over 2}+\nu} \tilde{\Psi}^{>\dagger}_{-{1\over 2}+\nu} |0,0;\nu\rangle
\end{equation}
on which $\mathrm{SU}(2)_{+}$ has an adjoint action. Thus, as in the case 
$\nu = {1\over 2}$, these states are space-time scalars that 
organize in the $\mathbf{1}\oplus \mathbf{3}$ of $\mathrm{SU}(2)_{+}$. 

The case in which $1/2 <\nu < 1$
is similar to the previous one. The four massless states are 
the matrix elements of 
\begin{equation}
\Psi^{<}_{{1\over 2}-\nu} 
\tilde{\Psi}^{<\dagger}_{{1\over 2}-\nu} |0,0;\nu\rangle .
\end{equation} 
These states are space-time scalars that 
organize in the $\mathbf{1}\oplus \mathbf{3}$ of $\mathrm{SU}(2)_{+}$.
\paragraph{The R-R twisted sectors}
There are neither bosonic nor fermionic zero modes in the orbifolded 
directions and the intercept is zero. The corresponding fields have a six 
dimensional dynamics and the massless fundamental is a bi-spinor of $SO(1,5)$.
The GSO projection selects the same chirality in the left and right 
sectors and  we obtain, as potential forms, a scalar and an antisymmetric 
self-dual two-form for a total of 1+3=4 states. These states are scalars of $SU(2)_{+}$. 
\subsection{Fractional D3-branes as sources}
\label{accoppia}
The boundary state describing the fractional D3-brane 
\cite{Diaconescu:2000dt,Billo:2000yb} associated to the
representation $\mathcal{D}_I$ is schematically given as 
follows:
\beq
\label{bs1}
\ket I = \mathcal{N}_3 \frac{d_I}{\sqrt{|\Gamma|}} \dket{a=0} +
\mathcal{N}^{\rm T}_3  \sum_{a\not=0}\sqrt{\frac{n_a}{|\Gamma|}}
\,\rho^a_I\,2\sin \pi\nu_{(a)} \dket{a}~.
\eeq
Here $\mathcal{N}_3$ is the usual normalization of the D3 boundary state,
$\mathcal{N}_3 = \sqrt{\pi}/2$, while the normalization in front of the twisted
components is~\cite{TAKA,Billo:2000yb}  $\mathcal{N}^{\rm T}_3 = \sqrt{\pi}
(2\pi\sqrt{\a'})^{-2}=\mathcal{N}_3/(2\pi^2\a')$
\footnote{Here we have restored the factor $2\pi\alpha'$
that we put equal 1 in the paper}. The Ishibashi states $\dket
a$ are in correspondence to the sectors of the closed string theory twisted by
an element of $\Gamma$ in the $a^{\rm th}$ conjugacy class and
$\nu_{(a)}$ defines the eigenvalues of the  2-dimensional representative of
such an element.
For the full expressions of the boundary states we refer to 
~\cite{Billo:2000yb}.
\par
The fractional D3 branes act as sources of closed string fields. It is possible
to deduce the linear couplings of a D-brane with the massless closed string 
states by simply projecting the latter  onto its boundary state 
\cite{DiVecchia:1997pr,DiVecchia:2000uf}. The untwisted
NS-NS component of the boundary state couples to the graviton $h_{\m\n}$, while
the R-R untwisted one couples to a 4-form potential $C_{(4)}$, and one finds:
\beq
\label{bs2}
{}_{\rm NS}\bra{I}h\rangle = -\frac{T_3\, d_IV_{4}}{\sqrt{|\Gamma|}} 
\sum_{\a=0}^3h_\a^{~\a}~,
\hskip 0.8cm
{}_{\rm R}\bra{I}{C_{4}}\rangle =\frac{\mu_3\, d_IV_{4}}{\sqrt{|\Gamma|}}\,C_{0123}~.
\eeq
Here $V_4$ is the world-volume of the
D3-brane, $T_3$ and $\mu_3$ are the usual
tension (in units of $\k$) and charge of a D3-brane, related to the
boundary state normalization by $\mu_3=\sqrt{2}T_3 = 2{\sqrt 2} \mathcal{N}_3$. 
As we said before,
in a non-trivially twisted NS-NS sector there are 4 massless scalar fields , 
organized in $\mathbf{1}+\mathbf{3}$ with respect to the residual 
SU$(2)$ geometrical symmetry. The  boundary state couples only to the singlet 
part, which we denote as $\tilde b_a$, as may be seen from the following discussion. The boundary conditions for world-sheet fermions may be written as follows
\begin{equation}
\Psi^{>}_{-n} = \ii \eta \tilde{\Psi}^{>}_{n}\,\, , \,
\Psi^{<}_{-m} = \ii \eta \tilde{\Psi}^{<}_{m}
\end{equation} 
where $n$, $m$ have the correct modings. Notice that one has
\begin{equation}
\psi^{1}_{-n}=\ii \eta \tilde{\psi}^{1}_{n}~,~~ 
\bar\psi^{1}_{-n}=\ii \eta \tilde{\bar\psi}^{1}_{n} .
\end{equation}
The oscillator part of the boundary state thus reads
\begin{equation}
\mathrm{exp}\left(\ii \eta \sum_{n<0} \Psi^{>\dagger}_{n}\tilde{\Psi}^{>}_{n}+ \ii \eta \sum_{m<0} \Psi^{<\dagger}_{m}\tilde{\Psi}^{<}_{m}\right) .
\end{equation}
The emission of massless states by a D3-brane, when $0<\nu<{1\over 2}$ for 
instance, is obtained by saturating the boundary state with the corresponding 
bra and is thus reads
\begin{eqnarray}
&&\langle0,0;\nu |\mathrm{Tr}(\sigma^{k} \Psi^{>}_{-{1\over 2}+\nu} 
\tilde{\Psi}^{>\dagger}_{-{1\over 2}+\nu})
\ee^{\ii \eta \sum_{n<0} \Psi^{>\dagger}_{n}\tilde{\Psi}^{>}_{n}+ \ii \eta
\sum_{m<0} \Psi^{<\dagger}_{m}\tilde{\Psi}^{<}_{m}} 
|0,0;\nu\rangle
\nonumber\\
&&= \mathrm{Tr}(\sigma^{k}\sigma^{0})
=\delta_{k,0}.
\end{eqnarray}
Hence only the state in the $\mathbf{1}$ of $SU(2)_{+}$ is emitted.

In the RR twisted sector, the
boundary state couples to a 4-form field $A^a_{4}$; explicitly, one has
\beq
\label{bs3}
{}_{\rm NS}\bra{I}\tilde b_a\rangle =-\frac{\mu_3\,V_4}{4\pi^2\alpha'}\,
\sum_{a\neq 0}\sqrt{\frac{n_a}{|\Gamma|}}
\,\rho^a_I\,2\sin \pi\nu_{(a)}
\,\tilde b_a~,
\eeq
\beq
\label{bs4}
{}_{\rm R}\bra{I}\tilde A^a_{4}\rangle= \frac{\mu_3\,V_4}{4\pi^2\alpha'}\,
\sum_{a\neq 0}\sqrt{\frac{n_a}{|\Gamma|}}
\,\rho^a_I\,2\sin \pi\nu_{(a)}A^a_{0123}~.
\eeq
If we limit ourselves to the case $I\neq 0$, corresponding to non-trivial representation of $\Gamma$, and use the following 
relation  \cite{Billo:2000yb} which connects the geometric twisted fields 
appearing in Section \ref{soluz} (that we indicate with $\Phi^i$) 
with the  ones  appearing in the present discussion
(${\Phi^a}$)
\beq
\label{bs5}
\Phi_i=-\sum_{a\neq 0}\sqrt{\frac{n_a}{|\Gamma|}}
\,\rho^a_l\,2\sin \pi\nu_{(a)}\frac{\Phi^a}{4 \pi^2\alpha'}
\eeq
both for $\Phi=b$ and $\Phi=A_4$, one can easily verify that
the couplings in \eq{bs2}-\eq{bs4} are perfectly consistent 
with those that one can read from the world-volume action
\eq{boundaryaction} after transforming the latter in terms of
canonically normalized fields. 
To this effect, one must take into account the fact that the tension and 
untwisted charge which one reads from the above boundary state 
are defined with respect to fields which are normalized correctly on the
covering space of the orbifold (namely, when in the action we integrate over 
the covering space). These correspond to $1/\sqrt{\Gamma|}$ times the
fields correctly normalized on the orbifold, which we used in \secn{soluz}. This
explains the further factor of $1/\sqrt{\Gamma|}$ present in the untwisted 
couplings in \eq{boundaryaction} with respect to \eq{bs2}. 
\end{document}